\documentclass[preprint,12pt]{revtex4-1}

\usepackage[brazil, english]{babel}
\usepackage{graphicx}
\usepackage{epsfig}
\usepackage{amssymb}
\usepackage{amsmath}
\usepackage{ulem}
\usepackage[T1]{fontenc} % if needed
\usepackage[utf8]{inputenc}
\usepackage{color}

\usepackage[titletoc]{appendix}

\begin{document}

\title{Magnetic catalysis and inverse magnetic catalysis in (2+1)-dimensional
    gauge theories from holographic models}
 
 \author{Diego M. Rodrigues$ ^{1} $}
\email[Eletronic address: ]{diegomr@if.ufrj.br}
\author{Eduardo Folco Capossoli$^{1,2,}$}
\email[Eletronic address: ]{educapossoli@if.ufrj.br}
\author{Henrique Boschi-Filho$^{1,}$}
\email[Eletronic address: ]{boschi@if.ufrj.br}  
\affiliation{$^1$Instituto de F\'\i sica, Universidade Federal do Rio de Janeiro, 21.941-972 - Rio de Janeiro-RJ - Brazil \\
 $^2$Departamento de F\'\i sica / Mestrado Profissional em Práticas da Educação Básica (MPPEB), Col\'egio Pedro II, 20.921-903 - Rio de Janeiro-RJ - Brazil}

\begin{abstract}
We study the deconfinement phase transition in $ (2+1) $-dimensional holographic $ SU(N) $ gauge theories in the presence of an external magnetic field from the holographic hard and soft wall models. We obtain exact solutions for the critical temperature of the deconfinement transition for any range of magnetic field. As a consequence, we find a critical magnetic field $(B_c)$, in which the critical temperature $(T_c)$ vanishes; for $B<B_c$ we have an inverse magnetic catalysis and for $ B>B_c $ we have a magnetic catalysis.
\end{abstract}

%\keywords{Holographic Models, (2+1) Gauge Theories, (Inverse) Magnetic Catalysis.}

\maketitle

\newcommand{\limit}[3]
{\ensuremath{\lim_{#1 \rightarrow #2} #3}}

\section{Introduction}   

The understanding of nonperturbative physics of Yang-Mills theory, especially QCD, remains an outstanding problem in modern theoretical physics. In particular, the effect of an external magnetic field in QCD has been the subject of many works \cite{Klimenko:1990rh, Klimenko:1991he, Klimenko:1992ch, Gusynin:1994re,Miransky:2002rp,Mizher:2010zb,Fraga:2008um,Gatto:2010pt,Gatto:2010qs,Osipov:2007je,Kashiwa:2011js,Alexandre:2000yf, Preis:2010cq, Ballon-Bayona:2013cta, Dudal:2015wfn, Mamo:2015dea, Li:2016gfn, Evans:2016jzo, Ballon-Bayona:2017dvv} over the years. Recently it has been observed in lattice QCD, in the context of chiral phase transition, an inverse magnetic catalysis (IMC), i.e., the decreasing of the critical temperature $(T_c)$ with increasing magnetic field $(B)$ for $eB \sim 1 $GeV$^2$ \cite{Bali:2011qj} and  more recently for $eB \sim 3 $GeV$^2$ \cite{Endrodi:2015oba}. This is in contrast with what would be expected: a magnetic catalysis (MC), meaning the increasing of the critical temperature with increasing magnetic field \cite{Andersen:2014xxa}. This behavior is also found in deconfinement phase transition. 

In this work, we study the deconfinement phase transition in $ (2+1) $-dimensional holographic pure $ SU(N) $ gauge theories, with $ N =2, 3$ and $N\rightarrow\infty $ in the presence of an external magnetic field. We restrict ourselves to $ (2+1) $ dimensions in order to simplify the analysis, guided by the fact that gauge theories in $ (2+1) $ dimensions are similar to the gauge theories in $ (3+1) $ dimensions \cite{Teper:1998te,Meyer:2003wx}. 
For instance, if we take a closer look (for a detailed account see \cite{Teper:1998te}), one can find that the perturbative sectors of these theories become free at high energies, the coupling sets the dynamical mass scale and becomes strong at small energies. Furthermore, they are linearly confining, just like in QCD in $ (3+1) $ dimensions for example. In addition, the lightest glueball state, $ 0^{++} $, has mass $m/\sqrt{\sigma}\approx 4$ in both (2+1)- and (3+1)-dimensional lattice gauge theories for any gauge group $SU(N)$. For these reasons we think that our work in $ (2+1) $ dimensions has significant theoretical interest for both lattice and gauge theories in general. Note, however, that in this work we are not dealing with the perturbative formulation of gauge theories but the nonperturbative approach based on the AdS/CFT correspondence or duality. 

A promising and fruitful approach to study IMC and MC is based on the AdS/CFT correspondence. Such formulation has a large use within strongly coupled gauge theories, including the nonperturbative IR physics of QCD \cite{Maldacena:1997re, Gubser:1998bc, Witten:1998qj, Witten:1998zw, Aharony:1999ti, Ramallo:2013bua}. However, in order to reproduce QCD physics in the IR region, one has to break the conformal invariance in the original AdS/CFT duality. The two most used models which realize this symmetry breaking are known as the hard and soft wall models, also known as AdS/QCD models \cite{Polchinski:2001tt, Polchinski:2002jw, BoschiFilho:2002ta, BoschiFilho:2002vd, deTeramond:2005su, BoschiFilho:2005yh, Karch:2006pv, Colangelo:2007pt, Capossoli:2013kb, Rodrigues:2016cdb}. In particular, some studies dealing with this framework discussed the IMC for small range of the magnetic field \cite{Dudal:2015wfn, Mamo:2015dea, Li:2016gfn, Evans:2016jzo}. However, these works could not predict what would happen for $eB>1 $GeV$^2$ because their approach is perturbative in the magnetic field $B$ and can only be trusted for weak fields $eB < 1 $GeV$^2$.

As a result of working in $(2+1)$ dimensions, here we obtain exact solutions for the critical temperature of the deconfinement transition for all range of the magnetic field. Surprisingly, and unexpectedly, we find a critical magnetic field $(B_c)$, in which the critical temperature $(T_c)$ vanishes, and for $B<B_c$ we find IMC and for $ B>B_c $ we find MC. Note that the choice to work in 2+1 dimensions implies that the unit of the magnetic field $B$ is given in terms of the string tension, throughout this text, instead the usual GeV scale. 

This work is organized as follows. In Sec. \ref{EMT}, after a brief reminder of the $ AdS_4/CFT_3 $ correspondence, we describe the geometric set up, equations of motion and its solutions for the models that we are going to consider in the remaining sections. In Sec. \ref{HW} we discuss the hard wall model, compute its on-shell Euclidean action (free energy) for thermal AdS and AdS-Black hole solutions and the free energy difference, which plays an important role in the deconfinement phase transition. In Sec. \ref{SW} we do the same analysis for the soft wall model. In Sec. \ref{PT} we study and present the results for the deconfinement phase transition for both hard and soft wall models. Finally, in Sec. \ref{Conc} we present our conclusions and final comments. We also include two appendices \ref{ApA} and \ref{ApB} showing in detail how to set the parameters used in each model. 

\section{Einstein-Maxwell Theory in (3+1) dimensions}\label{EMT}

In this section we review the Einstein-Maxwell theory in (3+1) dimensions together with the appropriate counterterms, which will be used to compute the finite on-shell actions in the next section.

Before presenting the Einstein-Maxwell theory in 4 dimensions, let us begin with a brief reminder of the basics concerning the AdS$ _4 $/CFT$ _3 $ Correspondence. The full gravitational background is the eleven-dimensional supergravity on $ AdS_4\times S^7 $. The dual field theory is the low-energy theory living on $ N$ $M2 $-branes on $ \mathbb{R}^{1,2} $, more specifically the $ \mathcal{N}=8 $ $ SU(N) $ Super-Yang-Mills theory in the large $ N $ limit \cite{Aharony:1999ti}. This theory has 8 supersymmetries and a global $ SO(8) $ R-symmetry group (the symmetry of the 7-sphere in the supergravity description). In the large $ N $ limit we have, on the gravity side, a classical supergravity on $ AdS_4\times S^7 $ while on the field theory side we have a strongly coupled (2+1)-dimensional gauge theory. 

Via Kaluza-Klein dimensional reduction, the supergravity theory on $ AdS_4\times S^7 $ may be consistently truncated to Einstein-Maxwell Theory on $ AdS_4$ \cite{Herzog:2007ij}. The action for this theory, in Euclidean signature \cite{comment1}, is given by: 
\begin{equation} \label{AdS4Action}
S = -\dfrac{1}{2\kappa^2_4}\int d^{4}x \sqrt{g}\left({R} -2\Lambda - L^{2}F_{\mu\nu}F^{\mu\nu}\right),
\end{equation}
where $ \kappa^2_4 $ is the 4-dimensional coupling constant, which is proportional to the 4-dimensional Newton's constant $( \kappa^2_4\equiv8\pi G_4 )$, $ d^{4}x\equiv d\tau\,dx_{1}\,dx_{2}\,dz $.  The Ricci scalar ${R} $ and the negative cosmological constant $ \Lambda $ for $ AdS_4$ are given, respectively,  by
\begin{eqnarray}
R &=& -\dfrac{12}{L^2}, \label{AdS4RicciSCALAR} \\
\Lambda &=& -\dfrac{3}{L^2}, \label{AdS4CosmologicalConst}
\end{eqnarray}
where $ L $ is the radius of $ AdS_4$. $ F_{\mu\nu} $ is the Maxwell field whose normalization comes from the reduction of the eleven-dimensional supergravity with coupling constant $ \kappa^2 $.  Furthermore, the coupling constants $ \kappa^2_4 $, for the large $N$ field theory, and $ \kappa^2 $ are related by:
\begin{eqnarray}
\dfrac{2L^2}{\kappa^2_4} &=& \dfrac{\sqrt{2}N^{3/2}}{6\pi}, \\
\dfrac{1}{2\kappa^2_4} &=& \dfrac{(2L)^{7}\mathrm{Vol}(S^7)}{2\kappa^2}.
\end{eqnarray}

According to the holographic renormalization \cite{Skenderis:2002wp}, the action \eqref{AdS4Action} must be regularized by adding a boundary counterterms. For our particular case, the boundary action is given by
\begin{equation}\label{AdS4Counter-term}
S_{Bndy} = - \dfrac{1}{\kappa^2_4}\int d^{3}x \sqrt{\gamma}\left(K + \dfrac{4}{L}\right),
\end{equation}
where $ \gamma $ is the determinant of the induced metric $\gamma_{\mu\nu}$ on the boundary, $ K = \gamma^{\mu\nu}K_{\mu\nu} $ is the trace of the extrinsic curvature $K_{\mu\nu} = -\frac{1}{2}(\nabla_{\mu}n_{\nu} + \nabla_{\nu}n_{\mu}) $ with $ n $ an outward-pointing normal vector to the boundary. The first term is just the Gibbons-Hawking surface term \cite{Gibbons:1976ue} in order to give a well-defined variational principle and the second one is a counterterms needed to cancel the UV divergences ($ z\rightarrow0 $) of the bulk action \eqref{AdS4Action}. For the Maxwell field, no new counterterms are needed since the action falls off sufficiently quickly near the boundary. Therefore, the renormalized action is given by subtracting a boundary term from the bulk action
\begin{equation}\label{AdS4ActionRenormalized}
S_{\mathrm{Ren}} = -\dfrac{1}{2\kappa^2_4}\int d^{4}x \sqrt{g}\left(R -2\Lambda - L^{2}F_{\mu\nu}F^{\mu\nu}\right) - \dfrac{1}{\kappa^2_4}\int d^{3}x \sqrt{\gamma}\left(K + \dfrac{4}{L}\right).
\end{equation}

The field equations coming from the bulk action \eqref{AdS4Action} together with the Bianchi identity are \cite{Herzog:2007ij}
\begin{eqnarray}
R_{MN} &=& 2L^2\left( F_{M}^{P}F_{NP}-\dfrac{1}{4}g_{MN}F^2\right)  - \dfrac{3}{L^2}g_{MN}, \label{FieldEquations}\\
\nabla_{M}F^{MN} &=& 0.
\end{eqnarray}
Our Ansatz for the metric and the background magnetic field to solve \eqref{FieldEquations} are given by
\begin{eqnarray}
ds^2 &=& \dfrac{L^2}{z^2}\left( f(z)d\tau^2 + \dfrac{dz^2}{f(z)} + dx^2_1 + dx^2_2\right), \label{AnsatzMetrica} \\
F &=& B\, dx_{1}\wedge dx_{2} \label{ansatzB}.
\end{eqnarray}
The background magnetic field $B$ that we introduce in this work belongs to a $ U(1) $ subgroup of $ SO(8) $ $ R $-symmetry group of the full theory. The behavior under a magnetic field is a classical probe of interacting (2+1)-dimensional systems \cite{Hartnoll:2007ai}. In order to understand the consequences of the bulk Maxwell field, let's consider the vector potential, which is a 1-form $ A $ such that $ F = dA $. So,
\begin{equation}
A = \dfrac{B}{2}(x_{1}dx_{2}-x_{2}dx_{1}).
\end{equation}
Therefore, one can note that the magnetic term remains finite at the $ AdS_4 $ boundary ($ z\rightarrow0 $). Thus, we can treat it as an external background magnetic field.

Using the Ans\"atze, \eqref{AnsatzMetrica} and \eqref{ansatzB}, the field equations \eqref{FieldEquations} are simplified and one has
\begin{eqnarray} 
z^2f''(z)-4zf'(z)+6f(z)-2B^{2}z^4-6 = 0, \label{FieldequationSimplified1} \\
zf'(z)-3f(z)-B^{2}z^4+3 = 0, \label{FieldequationSimplified2}
\end{eqnarray} 
with $ F^2 $ given by
\begin{equation} \label{MagneticField}
F^2 = \dfrac{2B^{2}z^4}{L^4}.
\end{equation}
Note that the Maxwell field $ F $ in \eqref{AdS4Action} and the magnetic field $ B $ have the same conformal dimension of (mass)$^2$ in 4 dimensions.

The two solutions of \eqref{FieldequationSimplified1} and \eqref{FieldequationSimplified2} which we are going to be interested are
\begin{eqnarray}
f_{Th}(z) &=& 1 + B^{2}z^4 \label{fT} \\
f_{BH}(z) &=& 1 + B^{2}z^3(z-z_H) - \dfrac{z^3}{z^{3}_H} \label{fBH}
\end{eqnarray} 
From the point of view of supergravity, the first solution, $ f_{Th}(z) $, corresponds to the thermal $ AdS_4 $, whereas the second solution, $ f_{BH}(z) $, corresponds to a black hole in $ AdS_4 $  where $ z_H $ is the horizon position, such that $ f_{BH}(z=z_H)=0 $. Both solutions are in the presence of an external background magnetic field $B$ and indeed satisfy the differential equations \eqref{FieldequationSimplified1} and \eqref{FieldequationSimplified2}. From the point of view of the boundary gauge theory, according to the holographic dictionary \cite{Witten:1998qj, Witten:1998zw}, the above solutions correspond to the gauge theory at zero and finite-temperature respectively, both in the presence of an external magnetic field, and have been found by the present authors recently in \cite{Rodrigues:2017cha}. It is important to mention that these solutions are exact in the magnetic field $B$ while in other references the corresponding solutions are perturbative in $B$, as one can see  in \cite{Dudal:2015wfn, Mamo:2015dea, Li:2016gfn}. 

\section{On-shell Euclidean Actions for the Hard Wall Model}\label{HW}

In this section, we compute the free energies from the on-shell Euclidean actions for the hard wall model using \eqref{AdS4ActionRenormalized}. In order to make it clear, we compute the action \eqref{AdS4Action} and the boundary action \eqref{AdS4Counter-term} separately for both thermal $ AdS_4 $ and $ AdS_4 $ black hole. Then we put those results together and calculate the free energy difference, $ \Delta S $, which will enable us to study the deconfinement phase transition in Sec. \ref{PT}.

The hard wall model \cite{Polchinski:2001tt,Polchinski:2002jw, BoschiFilho:2002ta,BoschiFilho:2002vd,BoschiFilho:2005yh,deTeramond:2005su} consists in introducing an IR hard cutoff $ z_{max} $ in the background geometry in order to break conformal invariance. The metric ansatz we use in this work is \eqref{AnsatzMetrica} in Euclidean signature with a compact time direction, $0\leq\tau\leq\beta$, with $ \beta = \frac{1}{T} $. The function $ f(z) $ is given by \eqref{fT} for the thermal $ AdS_4 $ and \eqref{fBH} for the $ AdS_4 $ black hole.
The introduction of a cutoff in this model means that 
\begin{equation}
0\leqslant z\leqslant z_{max},
\end{equation}
where $ z_{max} $ is the maximum value of the radial coordinate $ z $, and can be related to the mass scale of the boundary theory. For instance, in $ (3+1) $ dimensions $ z_{max} $ is usually related with energy scale of QCD \cite{BoschiFilho:2005yh,Capossoli:2013kb, Rodrigues:2016cdb} by,
\begin{equation}
z_{max} \sim \dfrac{1}{\Lambda_{QCD}}.
\end{equation}
Moreover we have to impose boundary conditions in $ z=z_{{max}} $. In this work, we use Neumann boundary conditions in order to fix $ z_{max} $ (for details see Appendix \ref{ApA}, where we also discuss the use of the Dirichlet boundary condition).

In the next two subsections, we compute the on-shell Euclidean actions for the thermal $ AdS_4 $ and the $ AdS_4 $ black hole, respectively.

\subsection{Thermal $ AdS_4 $}
\indent

Using the definitions of the Ricci scalar, the cosmological constant and the field strength given by \eqref{AdS4RicciSCALAR}, \eqref{AdS4CosmologicalConst} and \eqref{MagneticField} respectively, we can rewrite the action \eqref{AdS4Action} as
\begin{equation}
S = -\dfrac{1}{2\kappa^2_4}\int d^{4}x \sqrt{g}\left(\dfrac{6}{L^2} + \dfrac{2B^2z^4}{L^2}\right),
\end{equation}
with $ \sqrt{g} = \frac{L^4}{z^4} $. Therefore, the explicit on-shell Euclidean action for the thermal $ AdS_4 $ is computed as
\begin{equation}
S = - \dfrac{1}{2\kappa^2_4}\int_{0}^{\beta'} d\tau\, \mathcal{V}_2\,\int_{\epsilon}^{z_{max}} dz\, \dfrac{L^4}{z^4}\left(\dfrac{6}{L^2} + \dfrac{2B^2z^4}{L^2}\right),
\end{equation}
which gives
\begin{equation}
S = \dfrac{\beta'\mathcal{V}_{2}L^2}{\kappa^2_4}\left( \dfrac{1}{\epsilon ^3} - \dfrac{1}{z_{max}^3} \nonumber \\
+ B^{2}z_{max} + \mathcal{O}(\epsilon)\right) ,
\end{equation}
where $ \mathcal{V}_2\equiv \iint dx_{1}dx_{2} $,  $\epsilon$ is an UV cutoff which will be removed ($\epsilon \to 0$) at that end of calculations and $\beta'$ is the arbitrary period of the Euclidian time $\tau$ for the Thermal AdS$_4$ solution. 

For the boundary action \eqref{AdS4Counter-term}, we have
\begin{equation}
S_{Bndy} = \dfrac{\beta'\mathcal{V}_{2}L^2}{\kappa^2_4}\left(-\dfrac{1}{\epsilon^3} + \mathcal{O}(\epsilon)\right).
\end{equation}
Thus, the free energy for the thermal $ AdS_4 $, $ S_{Th} = S + S_{Bndy} $, is given by
\begin{equation}
S_{Th} = \dfrac{\beta'\mathcal{V}_{2}L^2}{\kappa^2_4}\left(-\frac{1}{z_{max}^3} + B^2z_{max} + \mathcal{O}(\epsilon)\right).
\end{equation}

\subsection{$ AdS_4 $ Black Hole}
\indent 

For the black hole case, we have
\begin{equation}
S' = - \dfrac{1}{2\kappa^2_4}\int_{0}^{\beta} d\tau\, \mathcal{V}_2\,\int_{\epsilon}^{z_{H}} dz\, \dfrac{L^4}{z^4}\left(\dfrac{6}{L^2} + \dfrac{2B^2z^4}{L^2}\right),
\end{equation}
which gives
\begin{equation}
S' = \dfrac{\beta\mathcal{V}_{2}L^2}{\kappa^2_4}\left( \dfrac{1}{\epsilon ^3} -\dfrac{1}{z_{H}^3} + B^{2}z_{H} +\mathcal{O}(\epsilon)\right),
\end{equation}
and for the boundary action \eqref{AdS4Counter-term}, we have
\begin{equation}
S'_{Bndy} = \dfrac{\beta\mathcal{V}_{2}L^2}{\kappa^2_4}\left(-\dfrac{1}{\epsilon^3} + \frac{1}{2z_{H}^3} + \dfrac{B^2 z_{H}}{2} + \mathcal{O}(\epsilon)  \right).
\end{equation}
Thus, the free energy for the $ AdS_4 $ black hole, $ S_{BH} = S' + S'_{Bndy} $, is given by
\begin{equation}
S_{BH} = \dfrac{\beta\mathcal{V}_{2}L^2}{\kappa^2_4}\left( -\dfrac{1}{2z_{H}^3} + \dfrac{3 B^2 z_{H}}{2} +\mathcal{O}(\epsilon)\right) .
\end{equation} 

\subsection{Hard Wall Free Energy Difference}

Now we define the general the free energy difference, $\Delta S$, given by
\begin{equation}\label{DeltaS_Definition}
\Delta S = \limit{\epsilon}{0}(S_{BH}-S_{Th}).
\end{equation}
Since we are comparing the two geometries at the same position $ z=\epsilon\rightarrow0 $ we can choose $ \beta' $ such that $ \beta' = \beta\sqrt{f(\epsilon)} = \beta $ \cite{Herzog:2006ra,Witten:1998zw}, since $ f(\epsilon) = 1 + \mathcal{O}(\epsilon^3) $ when $ \epsilon\rightarrow0 $, with $ f(z) $ given by \eqref{fBH}. Therefore, with this choice, we have that the free energy difference for the hard wall model is given by
\begin{equation}\label{FenergyDifHard}
\Delta S = \dfrac{\beta\mathcal{V}_{2}L^2}{\kappa^2_4}\left[\dfrac{1}{z_{max}^3} -\dfrac{1}{2z_{H}^3} + B^2 \left(\dfrac{3z_{H}}{2}-z_{max}\right)\right].
\end{equation}

Finally, it is important to mention that in the computations above we have assumed that $ z_{H}<z_{max} $. Otherwise we would have no transition at all, because the free energy difference would be a constant that never vanishes, given by
\begin{equation}
\Delta S = \dfrac{\beta\mathcal{V}_{2}L^2}{\kappa^2_4}\left( \frac{1}{2{z_H}^3} + \frac{B^2 z_H}{2}\right)  \qquad (z_{H}>z_{max})\,;
\end{equation}
which, in the limit $ B=0 $, gives
\begin{equation}
\Delta S = \dfrac{\beta\mathcal{V}_{2}L^2}{\kappa^2_4}\left( \frac{1}{2{z_H}^3}\right)\,,
\end{equation}
consistent with the higher-dimensional version of \cite{Herzog:2006ra}.

\section{On-shell Euclidean Actions for the Soft Wall Model}\label{SW}

In this section we compute the free energies for the soft wall model \cite{Colangelo:2007pt,Karch:2006pv}. The calculation is similar to the one described in the previous section. However, in this case we will have to introduce one more counterterm in the holographic renormalization scheme due to the introduction of a dilatonlike field.

For the soft wall model we consider the following 4-dimensional action
\begin{equation} \label{SoftwallAction}
S = -\dfrac{1}{2\kappa^2_4}\int d^{4}x \sqrt{g}\,e^{-\Phi(z)}\left({R} -2\Lambda - L^{2}F_{\mu\nu}F^{\mu\nu}\right),
\end{equation}
where $ \Phi(z) = kz^2 $ is the dilatonlike field, which has nontrivial expectation value and the constant $k$ is related to the QCD scale by $k\sim \Lambda_{QCD}^2$. In this work we are assuming that the dilaton field does not backreact on the background geometry. Moreover, as in \cite{Karch:2006pv}, we assume that our metric ansatz \eqref{AnsatzMetrica} satisfies the equations of motion for the full theory with $ f(z) $ given by \eqref{fT} for the thermal $ AdS_4 $ and \eqref{fBH} for the black hole in $ AdS_4 $. In \cite{Herzog:2006ra} it is argued that this should be the case because it conforms with the large $ N $ field theory expectations, at least qualitatively. 

In addition to the boundary action \eqref{AdS4Counter-term} we will have to include one more boundary action, $ S_{Bndy}^{\,\Phi} $, due to the dilatonlike field in this soft wall model, which will serve as a counterterm to cancel the bulk divergences. The simplest form for the boundary action which cancels this additional UV divergence in the soft wall model \eqref{SoftwallAction} is the following 
\begin{equation}
S_{Bndy}^{\,\Phi} = \dfrac{3}{\kappa^2_4}\int d^{3}x \sqrt{\gamma}\,\dfrac{\Phi}{L}.
\end{equation}
Therefore, the total boundary action for our 4-dimensional soft wall model \eqref{SoftwallAction}, $ S_{Bndy}^{\mathrm{Total}} $,  is given by
\begin{equation} \label{BndyActionSoft}
S_{Bndy}^{\mathrm{Total}} = - \dfrac{1}{\kappa^2_4}\int d^{3}x \sqrt{\gamma}\left(K + \dfrac{4}{L} - \dfrac{3\Phi}{L}\right).
\end{equation}

\subsection{Thermal $ AdS_4 $}
In this subsection we compute the free energy for the thermal $ AdS_4 $ for the soft wall model.

The calculation is similar to the one done in the previous section for the hard wall model, but in place of the action \eqref{AdS4Action} with a hard cutoff $ z_{max} $, we use \eqref{SoftwallAction}. Thus,

\begin{equation}
S = - \dfrac{1}{2\kappa^2_4}\int_{0}^{\beta'} d\tau\, \mathcal{V}_2\,\int_{\epsilon}^{\infty} dz\, \dfrac{L^4}{z^4}\,e^{-kz^2}\left(\dfrac{6}{L^2} + \dfrac{2B^2z^4}{L^2}\right),
\end{equation}
which gives
\begin{equation}
S = \dfrac{\beta'\mathcal{V}_{2}L^2}{\kappa^2_4}\left( \dfrac{1}{\epsilon^3} - \dfrac{3k}{\epsilon} + \dfrac{\sqrt{\pi }(B^2 + 4k^2)}{2 \sqrt{k}} + \mathcal{O}(\epsilon)\right).
\end{equation}

For the boundary action \eqref{BndyActionSoft} we have
\begin{equation}
S_{Bndy}^{\mathrm{Total}} = \dfrac{\beta'\mathcal{V}_{2}L^2}{\kappa^2_4}\left(-\dfrac{1}{\epsilon^3} + \dfrac{3 k}{\epsilon} + \mathcal{O}(\epsilon)\right). 
\end{equation}

Thus, the free energy for the thermal $ AdS_4 $, $ S_{Th} = S + S_{Bndy}^{\mathrm{Total}} $, in the soft wall model is given by
\begin{equation}
S_{Th} = \dfrac{\beta'\mathcal{V}_{2}L^2}{\kappa^2_4}\left(\dfrac{\sqrt{\pi }(B^2 + 4k^2)}{2 \sqrt{k}} + \mathcal{O}(\epsilon)\right).
\end{equation}

\subsection{$ AdS_4 $ Black Hole}
In this subsection, we compute the free energy for the $ AdS_4 $ black hole for the soft wall model.

Proceeding as was done in the previous section for the hard wall model we have
\begin{equation}
S' = - \dfrac{1}{2\kappa^2_4}\int_{0}^{\beta} d\tau\, \mathcal{V}_2\,\int_{\epsilon}^{z_{H}} dz\, \dfrac{L^4}{z^4}\,e^{-kz^2}\left(\dfrac{6}{L^2} + \dfrac{2B^2z^4}{L^2}\right),
\end{equation}
which gives
\begin{equation}
S' = \dfrac{\beta\mathcal{V}_{2}L^2}{\kappa^2_4}\left(  \dfrac{1}{\epsilon^3} -\dfrac{3 k}{\epsilon} + \dfrac{e^{-k z_{H}^2} \left(2 kz_{H}^2-1\right)}{z_{H}^3} + \dfrac{\sqrt{\pi}\left(B^2+4 k^2\right)\text{erf}(\sqrt{k}z_{H})}{2\sqrt{k}}\right),
\end{equation}
where \text{erf}$ (z) $ is the error function, defined as \text{erf}$ (z) $ = $ \frac{2}{\sqrt{\pi}}\int_{0}^{z}\,e^{-t^2}dt $.

For the boundary action \eqref{BndyActionSoft}, we have 
\begin{equation}
S_{Bndy}^{'\mathrm{Total}} = \dfrac{\beta\mathcal{V}_{2}L^2}{\kappa^2_4}\left(-\dfrac{1}{\epsilon^3} + \dfrac{3 k}{\epsilon} + \dfrac{1}{2 z_{H}^3} + \dfrac{B^2 z_{H}}{2}\right) . 
\end{equation}

Thus, the free energy for the $ AdS_4 $ black hole, $ S_{BH} = S' +  S_{Bndy}^{'\mathrm{Total}}$, in the soft wall model will be given by
\begin{eqnarray}
S_{BH} = \dfrac{\beta\mathcal{V}_{2}L^2}{\kappa^2_4} \bigg(&\dfrac{1}{2z_{H}^3}& + \dfrac{e^{-k z_{H}^2} \left(2kz_{H}^2-1\right)}{z_{H}^3} + \dfrac{B^2z_{H}}{2} + \nonumber \\
&+&\dfrac{\sqrt{\pi }\left(B^2+4 k^2\right) \text{erf}\left(\sqrt{k}z_{H}\right)}{2\sqrt{k}} + \mathcal{O}(\epsilon)\bigg).
\end{eqnarray}

\subsection{Soft Wall Free Energy Difference}

Here, we are going to compute the soft wall free energy difference $\Delta S$. 
Taking into account the same argument which led to $ \beta'=\beta\sqrt{f(\epsilon)} =\beta $ in the hard wall model, the free energy difference, $\Delta S= \displaystyle\lim_{\epsilon \to 0}(S_{BH}-S_{Th})$, for the soft wall model is given by
\begin{equation}\label{FenergyDifSoft}
\Delta S =  \dfrac{\beta\mathcal{V}_{2}L^2}{\kappa^2_4}\left(\dfrac{1}{2z_{H}^3} + \dfrac{e^{-kz_{H}^2} \left(2kz_{H}^2-1\right)}{z_{H}^3} + \dfrac{B^2z_{H}}{2} - \dfrac{\sqrt{\pi }\left(B^2+4 k^2\right)\text{erfc}\left(\sqrt{k}z_{H}\right)}{2\sqrt{k}}\right), 
\end{equation}
where erfc$ (z) $ is the complementary error function, defined as $\text{erfc}(z) = 1 - \text{erf}(z) $.

\section{Deconfinement Phase Transition}\label{PT}

In this section, we study the deconfinement phase transition of $ (2+1) $-dimensional gauge theories in the presence of a magnetic field for the hard and soft wall models. This transition  is a first order Hawking-Page phase transition 
 \cite{Hawking:1982dh, Witten:1998zw, Herzog:2006ra}. 
 To do so, we use the results we have found in the previous sections concerning the free energy differences imposing that this difference vanishes.

\subsection{Hard Wall Model}

In this subsection, we study the behavior of the critical temperature of deconfinement phase transition under an applied magnetic in $ (2+1) $-dimensional gauge theories for the hard wall model. In addition, we show the behavior of the critical horizon as a function of the applied magnetic field.

From the free energy difference for the hard wall model \eqref{FenergyDifHard} we have that 
\begin{equation}\label{FenergyDifHard2}
\Delta S(z_H,B;z_{max}) = \dfrac{\beta\mathcal{V}_{2}L^2}{\kappa^2_4}\left[\dfrac{1}{z_{max}^3} -\dfrac{1}{2z_{H}^3} + B^2 \left(\dfrac{3z_{H}}{2}-z_{max}\right)\right].
\end{equation}
One can note that for $ B = 0 $, we obtain the 3-dimensional version of \cite{Herzog:2006ra}:
\begin{equation}\label{FenergyDifHardB0}
\Delta S(z_H,B=0;z_{max}) = \dfrac{\beta\mathcal{V}_{2}L^2}{\kappa^2_4}\left(\dfrac{1}{z_{max}^3} -\dfrac{1}{2z_{H}^3}\right).
\end{equation}

We study the deconfinement phase transition by requiring
\begin{equation}\label{DeltaSB=0}
\Delta S(z_H=z_{H_C},B;z_{max}) = 0,
\end{equation}
where $ z_{H_C} $ is the critical horizon, from which we calculate the critical temperature through the formula
\begin{equation}\label{TcFormula}
T_c = \dfrac{|f'(z=z_{H_C})|}{4\pi},
\end{equation}
where $ f(z) $ is the horizon function given by \eqref{fBH}.

In the absence of magnetic field, we find, from \eqref{DeltaSB=0}, the following constraint equation for the critical horizon, $ z_{H_C} $,
\begin{equation}
\dfrac{1}{z_{max}^3} -\dfrac{1}{2z_{H_C}^3}=0.
\end{equation}
Thus, a phase transition occurs when $ z_{max}^3=2z_{H_C}^3 $ which, according to \eqref{TcFormula}, gives the critical temperature
\begin{equation}
T_c(B=0;z_{max}) = \frac{3}{2^{5/3}\ \pi z_{max} }\approx \dfrac{0.3}{z_{max}},
\end{equation}
which is the analogue of \cite{Herzog:2006ra,BallonBayona:2007vp} in $ (2+1) $ dimensions. Using the values of $ z_{max} $ obtained in the appendix \ref{ApA} we find that the critical temperature, $ T_c $, in units of the string tension, is given by
\begin{equation}\label{TcResultB=0Hard}
\dfrac{T_c(B=0)}{\sqrt{\sigma}}  = \left\{\begin{array}{ll} 0.45 \qquad SU(2); \\ 0.42 \qquad SU(3); \\
0.39 \qquad SU(N\rightarrow\infty).
\end{array}\right.
\end{equation} 
Therefore, for $ B=0 $, one can clearly see that the critical temperature of the deconfinement phase transition of $ SU(2) $ and $ SU(3) $ gauge theories in (2+1) dimensions is already close to the critical temperature of the large $N$ limit $ SU(N\rightarrow\infty) $ gauge theory, in agreement with the argument \cite{Athenodorou:2016ebg,Teper:1998te,Meyer:2003wx} that the physics of $ SU(N) $ gauge theories are close to $ N=\infty $ for $ N\geq2 $.

\begin{figure}
	\centering
	\includegraphics[scale = 0.4]{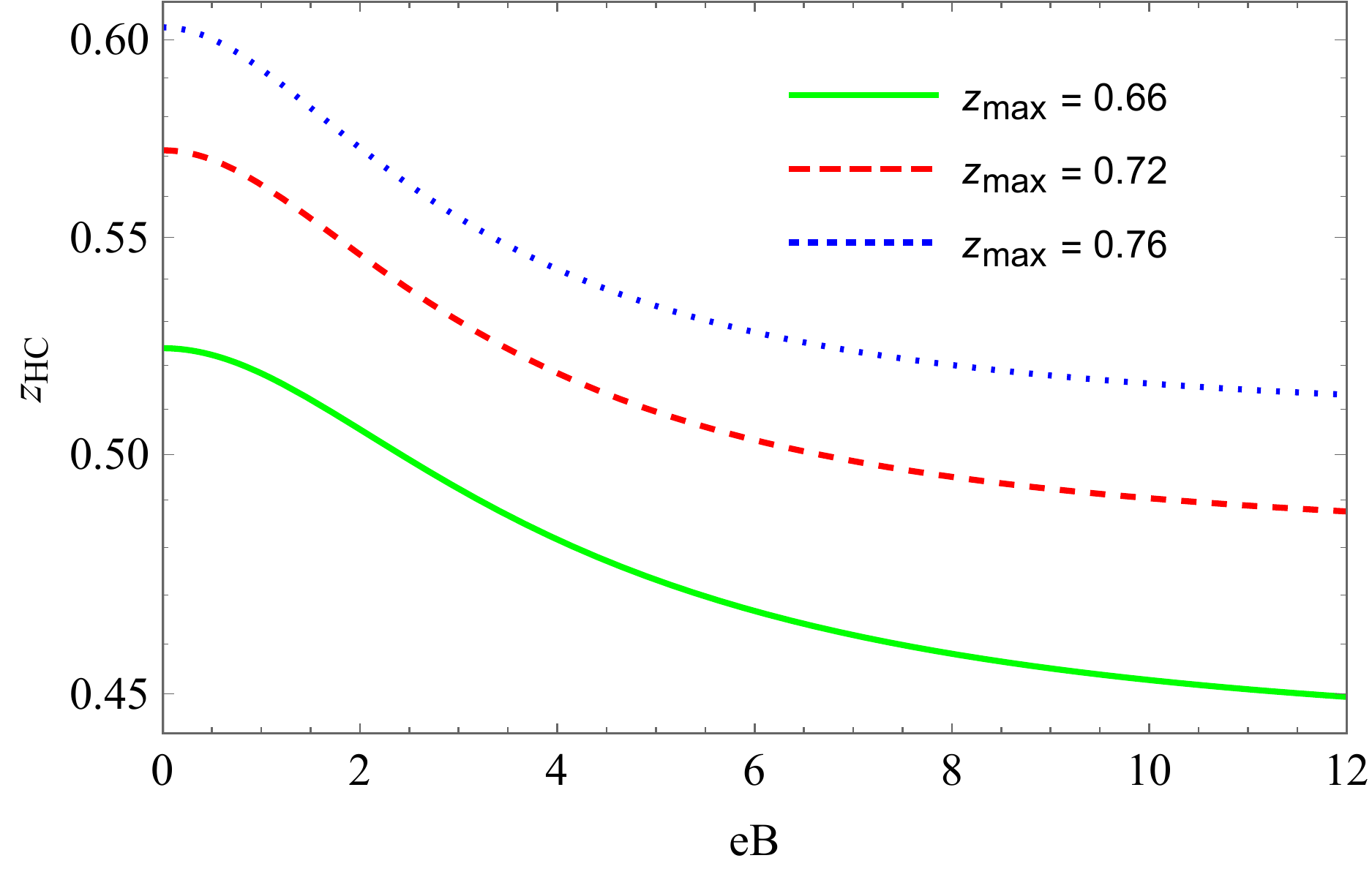}
	\caption{Critical horizon, $ z_{H_C}(B) $, as a function of the magnetic field, $ B $, for different values of the cutoff $ z_{max} $, corresponding to the $ SU(2) $, $ SU(3) $ and $ SU(N\rightarrow\infty) $ gauge theories in $ (2+1) $ dimensions, respectively, from the hard wall model. Here, we fixed the values of $ z_{max} $ using Neumann boundary conditions and lattice data \cite{Athenodorou:2016ebg}.}
	\label{fig:zhxbhardwall}
\end{figure}

For $ B\neq0 $, we have the following constraint equation for the critical horizon
\begin{equation}
\dfrac{1}{z_{max}^3} -\dfrac{1}{2z_{H_C}^3} + B^2 \left(\dfrac{3z_{H_C}}{2}-z_{max}\right)=0,
\end{equation}
which can be solved numerically for $ z_{H_C} \equiv z_{H_C}(B;z_{max}) $. The numerical result for $ z_{H_C} $ as a function of the magnetic field is shown in Figure \ref{fig:zhxbhardwall} for different values of $ z_{max} $, each one corresponding to a different gauge theory in $ (2+1) $ dimensions. One can see that the critical horizon decreases as we increase de magnetic field for the hard wall model. In the next subsection, for the soft wall model, it will be shown that $ z_{H_C} $ as a function of the magnetic field has the opposite behavior, it increases as we increase the magnetic field up to a certain value and then it saturates for higher magnetic fields.

\begin{figure}
	\centering
	\includegraphics[scale=0.4]{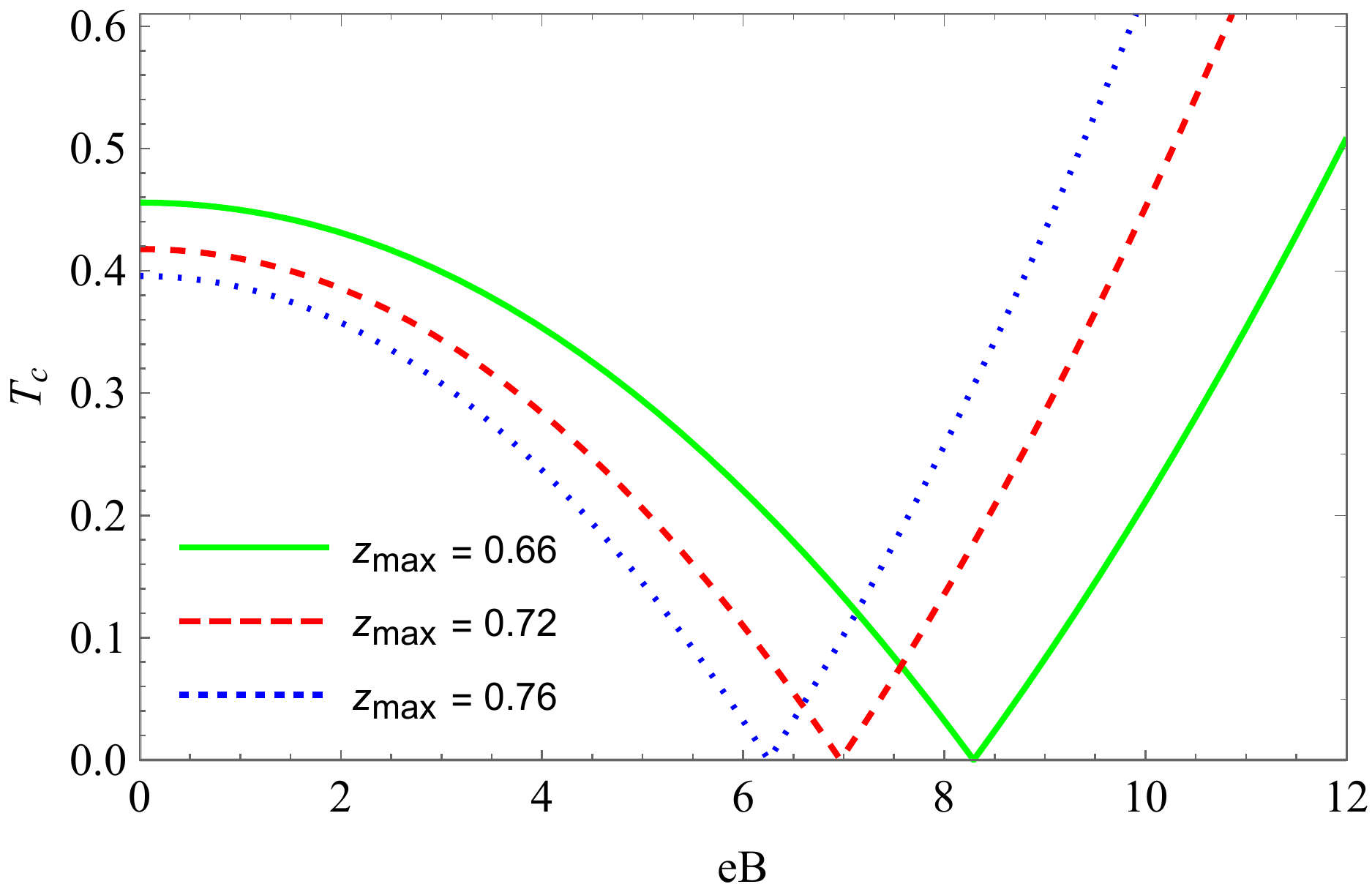}
	\caption{Critical temperature, $ T_{c}(B) $, as a function of the magnetic field, $ B $, for different values of the cutoff $ z_{max} $, corresponding to the $ SU(2) $, $ SU(3) $ and $ SU(N\rightarrow\infty) $ gauge theories in $ (2+1) $ dimensions, respectively, from the hard wall model. Here, we fixed the values of $ z_{max} $ using Neumann boundary conditions and lattice data \cite{Athenodorou:2016ebg}.}
	\label{fig:tcxbhardwall}
\end{figure}

Now, by using the black hole solution \eqref{fBH} and critical temperature $T_c$ given by \eqref{TcFormula}, one finds $T_c$ as a function of magnetic field $ B $:
\begin{equation}
T_{c}(B;z_{max}) = \dfrac{1}{4\pi}\left|-\dfrac{3}{z_{H_C}(B;z_{max})} + B^{2}z^3_{H_C}(B;z_{max}) \right|. 
\end{equation}
The corresponding numerical result for $T_{c}(B;z_{max})$ is shown in Figure \ref{fig:tcxbhardwall} for different values of the cutoff $ z_{max} $. One can note that our numerical results are consistent with our analytical results  \eqref{TcResultB=0Hard} for $ T_c(0) $. These numerical results are the exact solutions in the sense that they represent the behavior of the critical temperature as a function of the magnetic field. Note that this happens for any range of magnetic field. 
We chose this particular range in Figure \ref{fig:tcxbhardwall} to enhance the two phenomena we have found in this work. Concerning these two phenomena, one can see from Figure \ref{fig:tcxbhardwall} that we have a phase in which the critical temperature, $T_{c}(B)$, decreases with increasing magnetic field $B$, indicating an inverse magnetic catalysis (IMC). This phenomenon has been observed in lattice QCD  for $ eB\lesssim 1$ GeV$^2 $ \cite{Bali:2011qj} and  more recently for $eB \sim 3 $GeV$^2$ \cite{Endrodi:2015oba}. Since then many holographic approaches have studied  this behavior in $ (3+1) $ dimensions in both deconfinement and chiral phase transition contexts, see for instance \cite{Mamo:2015dea,Dudal:2015wfn,Evans:2016jzo,Li:2016gfn}. However, in many of these approaches just cited the problem could only be solved perturbatively in $B$, while in our results there is no restriction for the values or range of the magnetic field. Furthermore, we also predict a phase in which the critical temperature, $T_{c}(B)$, increases with increasing magnetic field $B$, indicating a magnetic catalysis (MC). This behavior was not found in these previous works cited above because, as was mentioned, the solution was valid only in a small range of magnetic field. Of course, these works tried to reproduce QCD in $ (3+1) $ dimensions, which is much more difficult than these QCD-like theories we are dealing with in one lower dimension.

The magnetic and inverse magnetic catalysis that we have found here for the hard wall model are separated by a critical magnetic field, $ B_c $. Note that the values of $ B_c $ depend on the gauge theory we are considering, which in turn depend on the cutoff $ z_{max} $. The values of the critical magnetic field found in this model, in units of the string tension squared, are the following
\begin{equation} \label{CriticalMagneticFieldHard}
\dfrac{B_c}{\sigma} = \left\{\begin{array}{ll} 8.29 \qquad SU(2); \\ 6.97 \qquad SU(3); \\
6.25 \qquad SU(N\rightarrow\infty).
\end{array}\right.
\end{equation}

\begin{figure}
	\centering
	\includegraphics[scale=0.4]{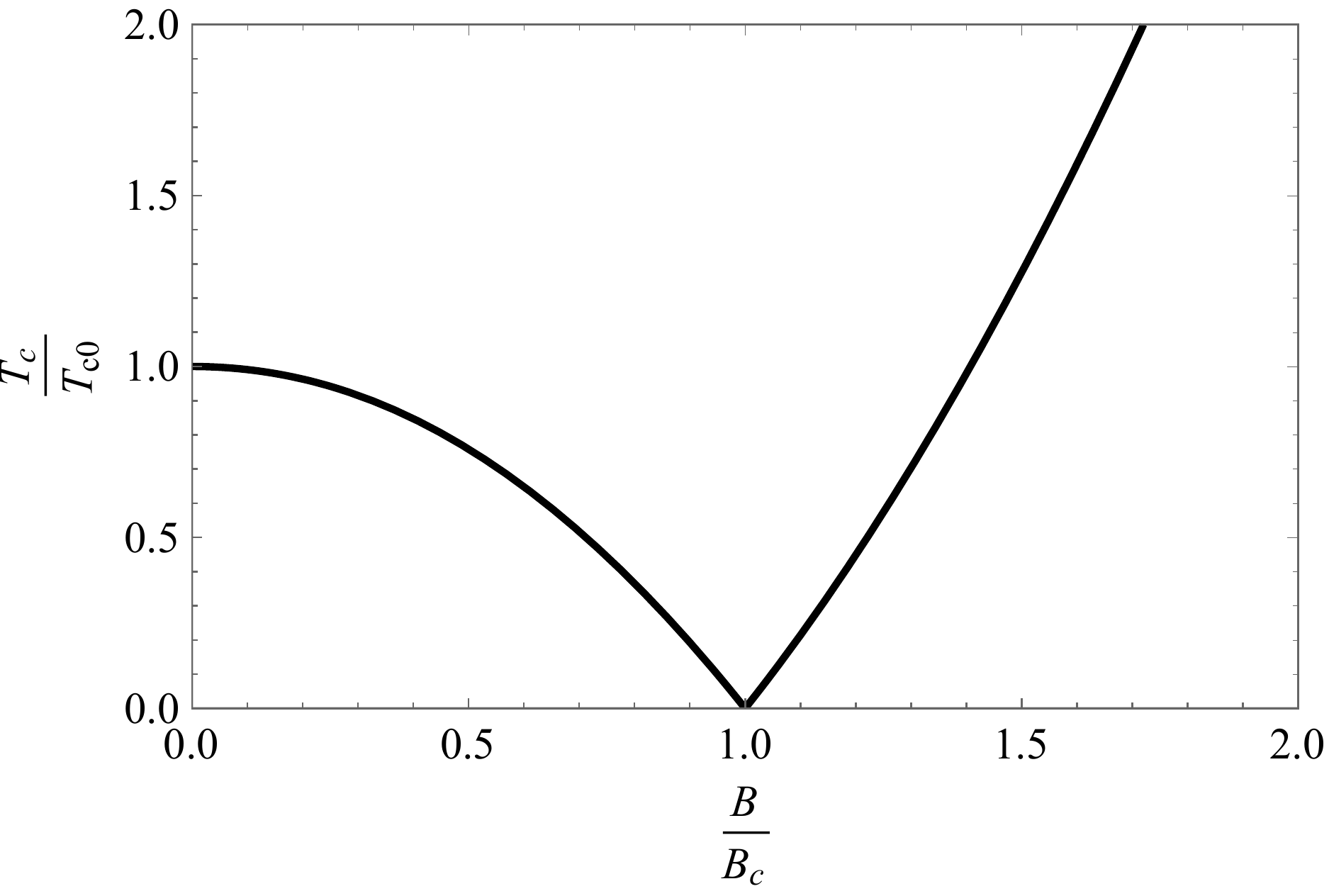}
	\caption{Normalized critical temperature, $ T_c/T_{c_0} $, as a function of $ B/B_c $ for the hard wall model for the three values of the cutoff $z_{max}$ corresponding to $SU(2)$, $SU(3)$, and $SU(N\to \infty)$. Note that the three corresponding curves coincide.}
	\label{fig:tcadimensionalxbadimensionalhardwall}
\end{figure}

Finally, in Figure \ref{fig:tcadimensionalxbadimensionalhardwall} we show the plot of the normalized critical temperature, $ T_c/T_{c_0} $, as a function of $ B/B_c $, where $ T_{c_0}\equiv T_c(B=0) $ and $ B_c $ is the critical magnetic field. 
Note that the values of the critical magnetic field are given by \eqref{CriticalMagneticFieldHard} for the $ SU(N) $ gauge theories in $ (2+1) $ dimensions with $ N = 2,3,\infty $ from the hard wall model.

\subsection{Soft wall Model}

In this subsection, we study the behavior of the critical temperature of deconfinement phase transition under an applied magnetic in $ (2+1) $-dimensional gauge theories from the soft wall model. In addition, we also show the behavior of the critical horizon as a function of the applied magnetic field.

The free energy difference for the soft wall model, \eqref{FenergyDifSoft}, explicitly reads
\begin{eqnarray}\label{FenergyDifSoft2}
\Delta S(z_H,B;z_{max}) =  \dfrac{\beta\mathcal{V}_{2}L^2}{\kappa^2_4}\bigg(&\dfrac{1}{2z_{H}^3}& + \dfrac{e^{-kz_{H}^2} \left(2kz_{H}^2-1\right)}{z_{H}^3} + \dfrac{B^2z_{H}}{2} \nonumber\\
&-& \dfrac{\sqrt{\pi }\left(B^2+4 k^2\right)\text{erfc}\left(\sqrt{k}z_{H}\right)}{2\sqrt{k}}\bigg). 
\end{eqnarray}
For $ B = 0 $, the condition for a phase transition requires
\begin{equation}
\dfrac{1}{z_{H_C}^3} + \dfrac{e^{-k z_{H_C}^2} \left(4 kz_{H_C}^2-2\right)}{z_{H_C}^3} - 4 \sqrt{\pi }k^{3/2}\text{erfc}\left(\sqrt{k}z_{H_C}\right) = 0.
\end{equation}
Thus, numerically there is a phase transition when $ \sqrt{k}z_{H_C} = 0.598671 $ which, after using \eqref{TcFormula}, gives the critical temperature 
\begin{equation}
T_{c}(B=0;k) = 0.397887 \sqrt{k},
\end{equation}
consistent with the treatment presented in \cite{Herzog:2006ra,BallonBayona:2007vp} for $ B=0 $ in one higher dimension. Now, using the values of $ k $ obtained in the appendix \ref{ApB} we find that the critical temperatures, $ T_c(0) $, in units of the string tension, for the $ SU(2) $, $ SU(3) $, and $ SU(N\rightarrow\infty) $ gauge theories in $ (2+1) $ dimensions, are given by
\begin{equation}\label{TcResultB=0Soft}
\dfrac{T_c(B=0)}{\sqrt{\sigma}}  = \left\{\begin{array}{ll} 0.77 \qquad SU(2); \\ 0.71 \qquad SU(3); \\
0.67 \qquad SU(N\rightarrow\infty).
\end{array}\right.
\end{equation}
As happened  for the hard wall model, the above results for the $ SU(2) $ and $ SU(3) $ gauge theories in $ (2+1) $ dimensions are close to $ N=\infty $, again in agreement with \cite{Athenodorou:2016ebg,Teper:1998te}. 

\begin{figure}
	\centering
	\includegraphics[scale=0.49]{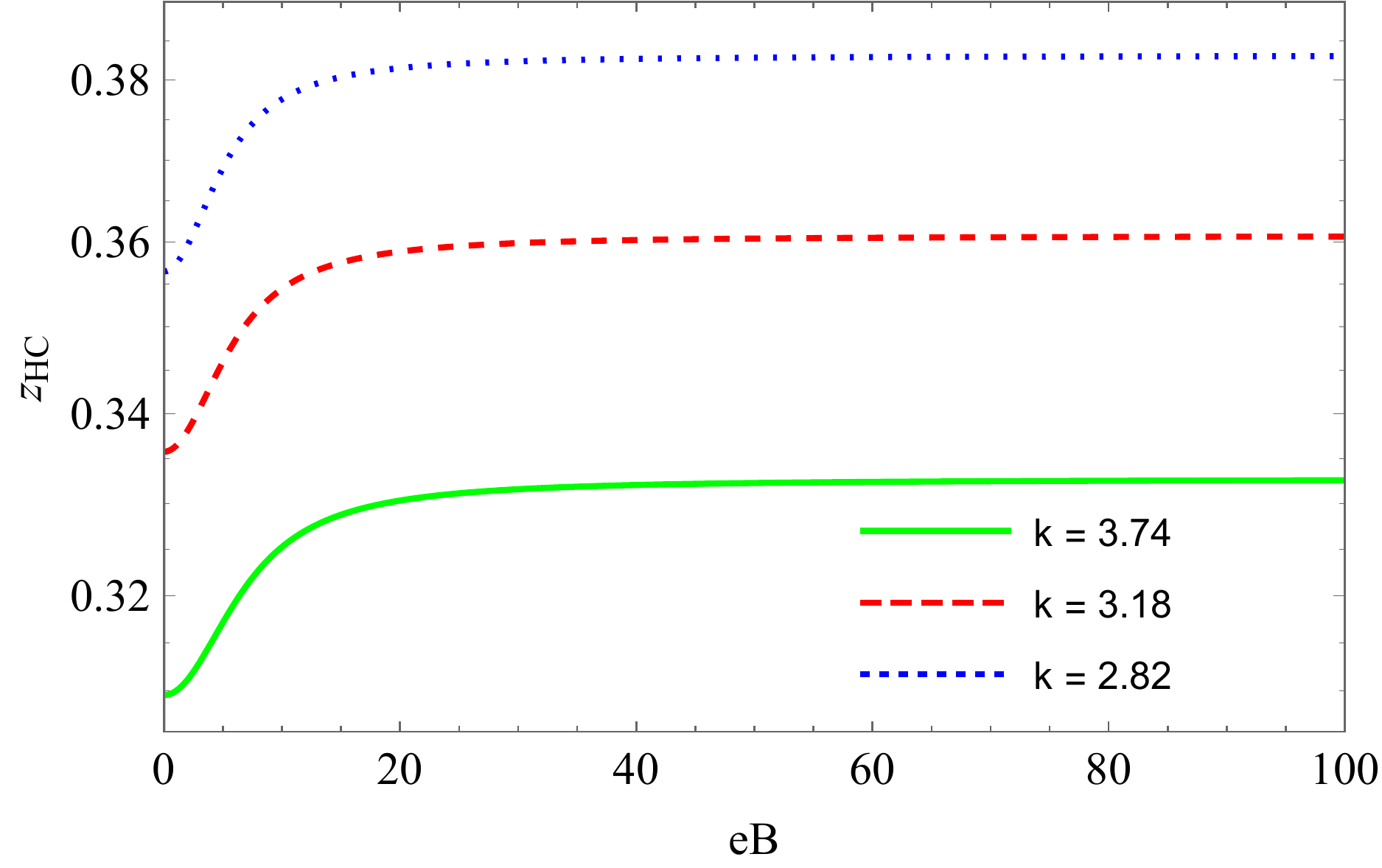}
	\caption{Critical horizon, $ z_{H_C}(B) $, as a function of the magnetic field, $ B $, for different values of the dilaton constant $ k $, corresponding to the $ SU(2) $, $ SU(3) $ and $ SU(N\rightarrow\infty) $ gauge theories in $ (2+1) $ dimensions, respectively from the soft wall model. Here, we fixed the values of $ k $ from lattice data \cite{Athenodorou:2016ebg}.}
	\label{fig:zhxbsoftwall}
\end{figure}
For $ B\neq0 $, we have the following constraint for the critical horizon, $ z_{H_C} $,
\begin{equation}
\left(\dfrac{1}{2z_{H}^3} + \dfrac{e^{-kz_{H}^2} \left(2kz_{H}^2-1\right)}{z_{H}^3} + \dfrac{B^2z_{H}}{2} - \dfrac{\sqrt{\pi }\left(B^2+4 k^2\right)\text{erfc}\left(\sqrt{k}z_{H}\right)}{2\sqrt{k}}\right) = 0,
\end{equation}
which can be solved numerically for $ z_{H_C} $. The numerical result for $ z_{H_C} $ as a function of the magnetic field is shown in Figure \ref{fig:zhxbsoftwall} for different values of $ k $, each one corresponding a different gauge theory in $ (2+1) $ dimensions. One can see, in contrast with the hard wall model discussed in the previous subsection, that the critical horizon, $ z_{H_C}(B) $, increases with increasing magnetic field up to a certain point and then saturates for higher magnetic fields in the soft wall model. 

\begin{figure}
	\centering
	\includegraphics[scale=0.49]{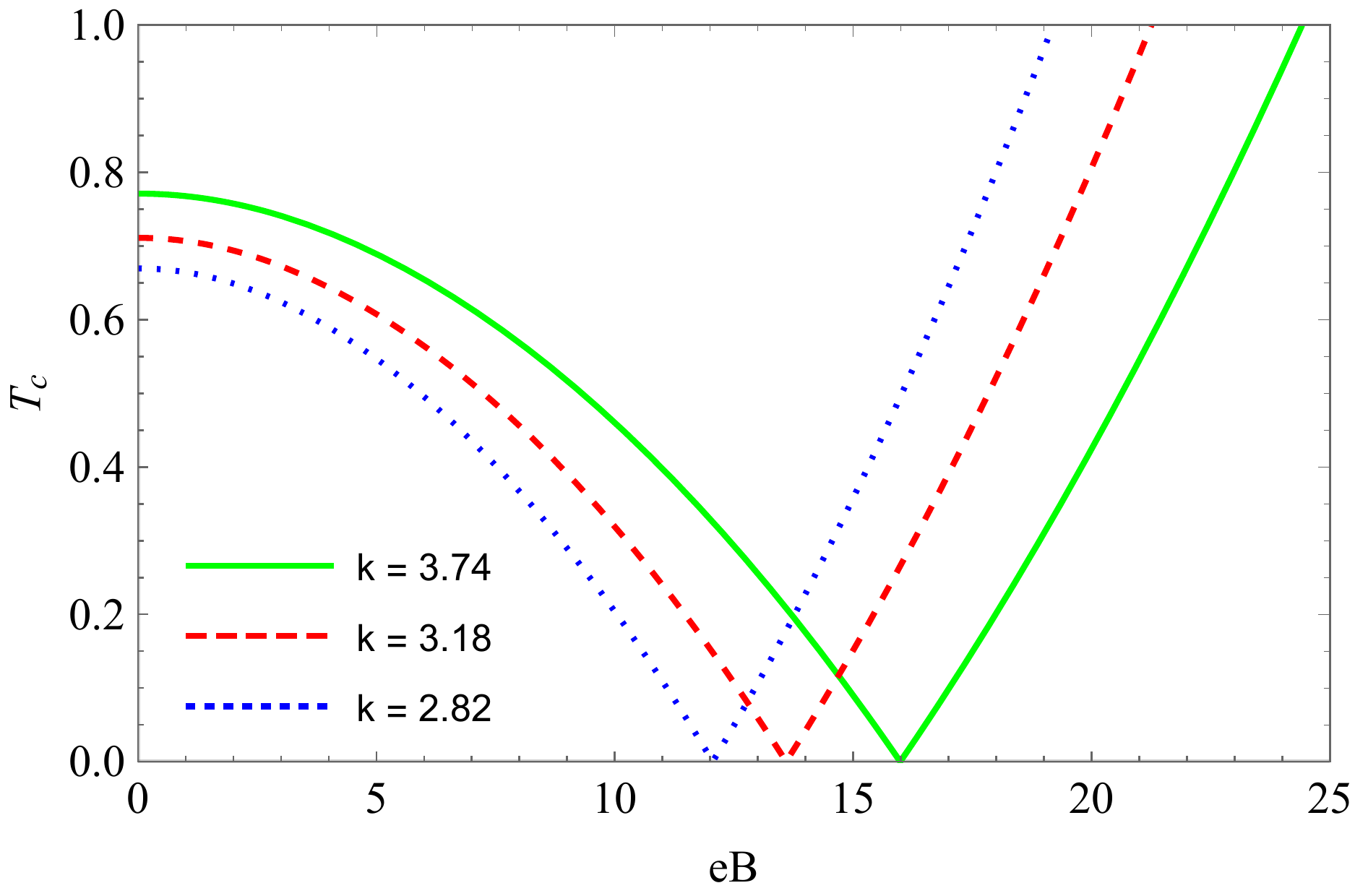}
	\caption{Critical temperature, $ T_{c}(B) $, as a function of the magnetic field, $ B $, for different values of the dilaton constant $ k $, corresponding to the $ SU(2) $, $ SU(3) $ and $ SU(N\rightarrow\infty) $ gauge theories in $ (2+1) $ dimensions, respectively from the soft wall model. Here, we fixed the values of $ k $ from lattice data \cite{Athenodorou:2016ebg}.}
	\label{fig:tcxbsoftwall}
\end{figure}

For the critical temperature as a function of the magnetic field, $T_{c}(B)$, the numerical result is shown in Figure \ref{fig:tcxbsoftwall} for different values of $ k $. Here the numerical results are consistent with our analytical results  \eqref{TcResultB=0Soft} for $T_c(0)$. As in the hard wall model, these numerical results are the exact solutions in the sense that they represent the behavior of the critical temperature as a function of the magnetic field for any range of magnetic field. As one can note in Figure \ref{fig:tcxbsoftwall} we also obtained the magnetic and inverse magnetic catalysis phases, separated by a critical magnetic field, whose values are larger than those obtained in the hard wall model. For the soft wall model, the critical magnetic field depends on $ k $, each one corresponding to a different gauge theory on the boundary. The values of the critical magnetic field found in this model, in units of the string tension squared, are the following
\begin{equation} \label{CriticalMagneticFieldSoft}
\dfrac{B_c}{\sigma} = \left\{\begin{array}{ll} 15.9 \qquad SU(2); \\ 13.6 \qquad SU(3); \\
12.1 \qquad SU(N\rightarrow\infty).
\end{array}\right.
\end{equation}

\begin{figure}
	\centering
	\includegraphics[scale=0.49]{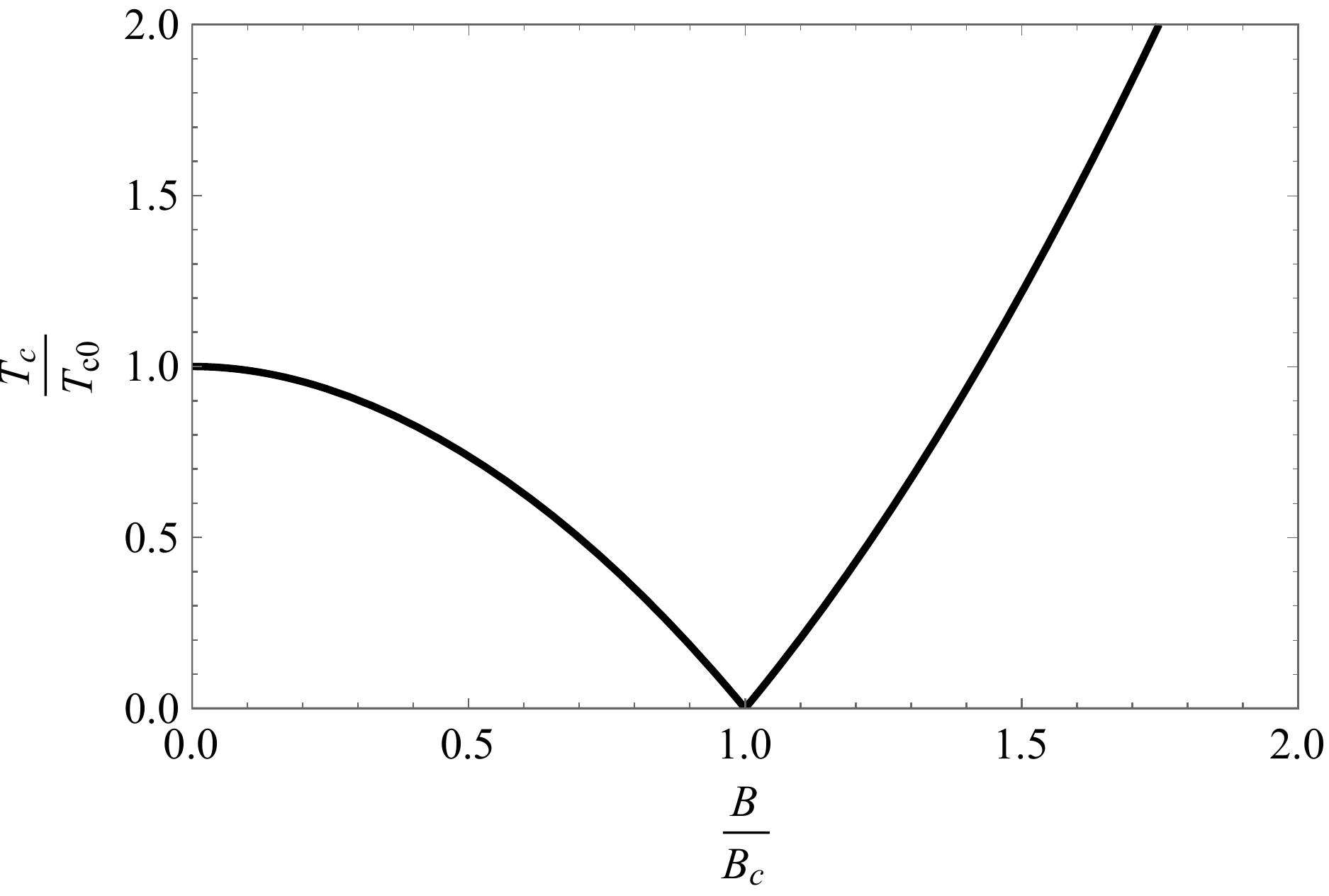}
	\caption{Normalized critical temperature, $ T_c/T_{c_0} $, as a function of $ B/B_c $, for the soft wall model with different values of $k$ for the gauge groups $SU(2)$, $SU(3)$, and $SU(N\to \infty)$, as discussed in appendix \ref{ApB}. Note that the three corresponding curves coincide.}
	\label{fig:tcadimensionalxbadimensionalsoftwall}
\end{figure}

In Figure \ref{fig:tcadimensionalxbadimensionalsoftwall} we show the plot of the normalized critical temperature, $ T_c/T_{c_0} $, as a function of $ B/B_c $, where $ T_{c_0}\equiv T_c(B=0) $ and $ B_c $ is the critical magnetic field. The values of  $ B_c $ are given by \eqref{CriticalMagneticFieldSoft} for the $ SU(N) $ gauge theories in $ (2+1) $ dimensions with $ N = 2,3,\infty $ from the holographic soft wall model. 

\begin{figure}
	\centering
	\includegraphics[scale=0.45]{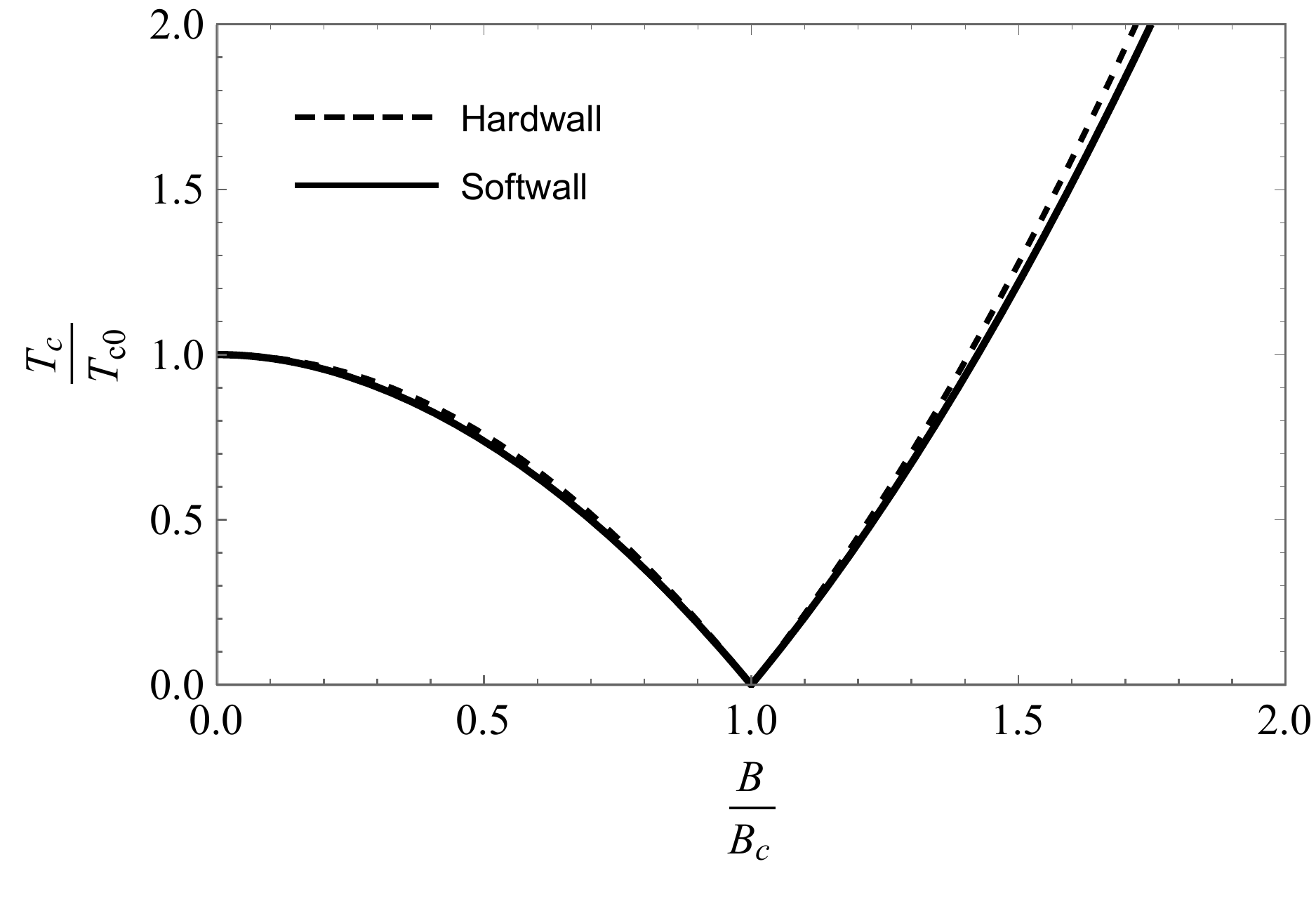}
	\caption{The ratio of the critical temperatures as a function of the ratio of the magnetic fields for the hard and soft wall models.}
	\label{HardSoft}
\end{figure}

Finally, for comparison, we plot in Figure \ref{HardSoft} the normalized critical temperature, $ T_c/T_{c_0} $, as a function of $ B/B_c $, for both hard and soft wall models. From this figure we can clearly see that the predictions for the two models are very similar.

\newpage
\section{Conclusion and Discussion}\label{Conc}

In this work we studied the problem of the deconfinement phase transition in the presence of an external magnetic field in $SU(N)$ gauge theories in $ (2+1) $ dimensions using two different holographic models. This work is a more detailed version of  \cite{Rodrigues:2017cha}. 
This study was motivated by recent lattice results indicating inverse magnetic catalysis in the context of chiral phase transition \cite{Bali:2011qj}. Previous studies include \cite{Gusynin:1994re,Miransky:2002rp,Mizher:2010zb,Fraga:2008um,Gatto:2010pt,Gatto:2010qs,Osipov:2007je,Kashiwa:2011js,Alexandre:2000yf, Preis:2010cq, Ballon-Bayona:2013cta}. 
Since then, many different works appeared dealing with IMC in both chiral and deconfinement phase transitions within holographic models \cite{Mamo:2015dea,Dudal:2015wfn,Evans:2016jzo,Li:2016gfn}. However, these works were valid only for a small range of the magnetic field. Here we obtained the solution for any magnetic field. Then, we were able to find the IMC and MC phases separated by a critical magnetic field $B_c$ in (2+1) dimensions. 

Since we have worked in $ (2+1) $ dimensions, physical quantities such as the critical temperature, $T_c$, the magnetic field, $B$, and the critical magnetic field, $B_c$, are not measured in GeV or MeV. However, one might consider gauge theories  in lower-dimensional condensed matter physics setup as effective field theories where one can bring in physical units. Anyway, in place of the string tension, $ \sqrt{\sigma} $, frequently used in lattice simulations \cite{Athenodorou:2016ebg,Teper:1998te,Meyer:2003wx}, the most natural physical unit we could have used in this work is the gauge theory coupling constant, $ g^2 $, which has dimensions of mass in $ (2+1) $ dimensions. Moreover, the fact that we are in $ (2+1) $ dimensions make things easier also for lattice calculations, which are much more accurate and the computational cost is much smaller than in $ (3+1) $ dimensions \cite{Teper:1998te}. Furthermore, since the large $ N $ limit is much simpler than the physically interesting $ N=2,3 $ theory (depending on whether we are in 3 or 4 dimensions), it is a relevant problem to study the physics in the large $ N $ approximation in order to get a better understanding of the $ N=2,3 $ theory in 3, 4 dimensions. In fact, we have seen that the critical temperature  $T_c$ for the $ SU(2) $ gauge theory is already close to the critical temperature for the $ SU(N\rightarrow\infty) $ gauge theory, in agreement with lattice results both in $ (2+1) $ and $ (3+1) $ dimensions.  

Recently in \cite{Giataganas:2017koz}, it was found that the IMC could be explained by anisotropy (with no magnetic field) in $(3+1)$ dimensions. For large fields, they also found that there might be a competing effect due to B yielding the MC. 

Currently, we are investigating the chiral phase transition and symmetry restoration in the presence of an external magnetic field in $ (2+1) $ dimensions from holographic models \cite{Diego_CPhT}, inspired by many recent works in $ (3+1) $ dimensions, including especially \cite{Chelabi:2015cwn, Chelabi:2015gpc, Li:2016gfn}.

\vspace{12pt}
\noindent {\bf Acknowledgments:} 
We would like to thank Luiz F. Ferreira, Adriana Lizeth Vela, Renato Critelli, Rômulo Rogeumont, and Marco Moriconi for helpful discussions during the course of this work. We also thank Elvis do Amaral for the help with numerical solutions. We would also like to thank Michael Teper,  Juan F. Pedraza,   Konstantin Klimenko and Gergely Endrodi for useful correspondence. The calculations were done using the MATHEMATICA package diffgeo. D.M.R is supported by Conselho Nacional de Desenvolvimento Científico e Tecnológico (CNPq), E.F.C. is partially supported by PROPGPEC-Colégio Pedro II, and H.B.-F. is partially supported by CNPq.

\appendix

\section{Fixing The Cutoff $ z_{max} $ in the Hard Wall Model}\label{ApA}

In this appendix, we construct a holographic picture (in 4 dimensions) for glueball spectra in 3 dimensions based on the higher-dimensional version \cite{BoschiFilho:2005yh,Rodrigues:2016cdb} in order to fix the slice $ z_{max} $ in the $ AdS_4 $ space. In \cite{Herzog:2006ra}, $ z_{max} $ was associated with the lightest $ \rho $ meson mass. In this work, we follow a similar method, but, as it will be clear in the end of this construction, we will associate $ z_{max} $ with the lightest glueball state in (2+1) dimensions with gauge groups $ SU(2) $, and $ SU(3) $ as well as the large $ N $ limit $ SU(N\rightarrow\infty) $ from the lattice \cite{Athenodorou:2016ebg,Teper:1998te, Meyer:2003wx}. 

First, let us consider the equation of motion for a scalar field with mass $M_{4}$ in $ AdS_{4} $ 
\begin{equation} \label{AdS scalar eq. motion}
\left[z^2\partial_z \dfrac{1}{z^2}\partial_z + \eta^{\mu\nu}\partial_\mu\partial_\nu -\dfrac{(M_{4}L)^2}{z^2}\right] \Phi(z,x^\mu) = 0,
\end{equation}
where $ \eta^{\mu\nu} = diag(-,+,+)$ is the Minkowski metric in (2+1) dimensions and $ L $ is the radius of $ AdS_4 $. From the the AdS/CFT dictionary \cite{Ramallo:2013bua} we have, in $ d+1 $ dimensions,
\begin{equation}
(M_{d+1}L)^2 = \Delta(\Delta-d),
\end{equation}
where $ \Delta $ is the conformal dimension, which can be written as
\begin{equation}
\Delta = \dfrac{d}{2} + \nu,
\end{equation}
where $ \nu = \sqrt{\frac{d^2}{4}+(M_{d+1}L)^2} $. Thus, in $ d=3 $ dimensions, we have
\begin{equation}\label{massrelation3D}
(M_{4}L)^2 = \Delta(\Delta-3),
\end{equation}
with 
\begin{equation}
\Delta = \dfrac{3}{2} + \nu; \quad \nu = \sqrt{\dfrac{9}{4}+(M_{4}L)^2}  .
\end{equation}
Using the Ansatz $ \Phi(z,x^\mu) = e^{-iP_{\mu}x^\mu}\, \phi(z) $, where $ P^2 = - m^2 $ with $ m $ being the mass of glueball states in 3 dimensions, we can write \eqref{AdS scalar eq. motion} as
\begin{equation}
z^2\phi''(z)-2z\phi'(z)+((mz)^2-(M_{4}L)^2)\phi(z)=0.
\end{equation}
This is the Bessel equation and the solutions are given by
\begin{equation}
\phi(z) = z^{3/2}\left[c_1\,J_{\nu}(m_{\nu,k}\,z) +c_2\,N_{\nu}(m_{\nu,k}\,z)\right],
\end{equation}
where $ c_1 $ and $ c_2 $ are normalization constants, $J_\nu(z) $ and $N_\nu(z) $ are the Bessel and Neumann functions, respectively. Since we are interested in regular solutions inside $ AdS_4 $ space, we are going to disregard the Neumann solution because it will act as a source for the field theory operator on the boundary, according to Witten's prescription \cite{Witten:1998qj}. 
In this particular case, the physical solution for $ \Phi(z,x^\mu) $ becomes: 
\begin{equation}
\Phi_{\nu,k}(z,x^\mu) = C_{\nu,k}\,e^{-iP_{\mu}x^{\mu}}\,z^{3/2}\,J_{\nu}(m_{\nu,k}\,z), 
\end{equation}
where $ \nu = \sqrt{\frac{9}{4}+(M_{4}L)^2} $, $ m_{\nu,k} $ are the masses of glueball states and $ k=2,3,..., $ represents the radial excitations, with $ k=1 $ being the ground state. Since we are going to focus only on the ground state mass we will omit the index $ k $ from now on.

Now we must impose boundary conditions at $ z=z_{max} $. For Dirichlet boundary conditions,
\begin{equation}
\Phi(z=z_{max},x^\mu)=0,
\end{equation}
we have
\begin{equation}\label{Dboundarycondition}
J_{\nu}(m_{\nu}\,z_{max})=0,
\end{equation}
and for Neumann boundary conditions,
\begin{equation}
\partial_{z}\Phi(z=z_{max},x^\mu)=0,
\end{equation}
we have
\begin{equation}\label{Nboundarycondition}
2 z_{max} J_{\nu-1}(m_{\nu}\,z_{max}) + (3-2\nu)J_\nu(m_{\nu}\,z_{max}) = 0
\end{equation}

Following \cite{deTeramond:2005su} we are going to associate $ \nu $ with the spin $ J $ of the glueball states. To see this consider the glueball operator $ \mathcal{O} = F^2 $, where $ F $ is the field strength. In 3 dimensions this operator has conformal dimension $ \Delta = 3 $. Now consider the same operator $ \mathcal{O} $ with $ J $ insertions of covariant derivatives $ D_{\mu} $
\begin{equation}
\mathcal{O}_{J} = FD_{\{\mu_1...}D_{\mu_J\}} F.
\end{equation}
In this case the operator $ \mathcal{O}_{J} $ has dimensions $ \Delta = 3+J $. Therefore the mass relation becomes:
\begin{equation}\label{massrelation3D2}
(M_{4}L)^2 = J(J+3).
\end{equation}
Using this relation in the equation for $ \nu $, we have
\begin{equation}
\nu = \dfrac{1}{2}(3+2J).
\end{equation}
So, for the scalar glueball case, $ J=0 $, we have from \eqref{Dboundarycondition} and \eqref{Nboundarycondition}, respectively
\begin{eqnarray}\label{boundarycondition2}
J_{3/2}(m_{3/2}\,z_{max})&=&0; \quad \text{(Dirichlet b.c.)}\\
J_{1/2}(m_{1/2}\,z_{max})&=&0; \quad \text{(Neumann b.c.)}
\end{eqnarray}
Therefore, the ground state mass for the scalar glueball is given by the first zero of $ J_{3/2}(\lambda) $ or $ J_{1/2}(\lambda) $, which are $ \lambda = 4.493 $ and $ \lambda = 3.141 $, respectively. Thus, we have
\begin{eqnarray} 
m_{3/2} &=& \dfrac{4.4934}{z_{max}}, \label{DFixZmax} \\ 
\vspace{1pt}
m_{1/2} &=& \dfrac{3.141}{z_{max}} \label{NFixZmax}.
\end{eqnarray}
From a recent lattice result in (2+1) dimensions \cite{Athenodorou:2016ebg}, the mass of the lightest glueball state, $ 0^{++} $, in units of string tension and in the continuum limit for the gauge group $ SU(2) $, $ SU(3) $ and the large $ N $ limit $ SU(N\rightarrow\infty) $, is given by
\begin{equation}\label{Masses_Lattice}
\dfrac{m_{0^{++}}}{\sqrt{\sigma}} = \left\{\begin{array}{lll} 4.7367 \qquad SU(2), \\ 4.3683\qquad SU(3),\\
4.116 \qquad SU(N\rightarrow\infty),
\end{array}\right.
\end{equation}
where $ \sqrt{\sigma} $ is the string tension \cite{Teper:1998te}. 

Therefore, using \eqref{DFixZmax} for Dirichlet boundary conditions we can fix $ z_{max} $ as
\begin{equation}
z_{max}\sqrt{\sigma} = \left\{\begin{array}{ll} 0.949 \qquad SU(2), \\ 1.03 \qquad SU(3), \\
1.09 \qquad SU(N\rightarrow\infty),
\end{array}\right.
\end{equation}
and using \eqref{NFixZmax} for Neumann boundary conditions, we have
\begin{equation}
z_{max}\sqrt{\sigma} = \left\{\begin{array}{ll} 0.66 \qquad SU(2), \\ 0.72 \qquad SU(3), \\
0.76 \qquad SU(N\rightarrow\infty),
\end{array}\right.
\end{equation}
which is, as expected, expressed in units of the inverse of the string tension.

\section{Glueball Spectra in $ d $ dimensions and $ k $ fixing in the Soft Wall Model}\label{ApB}

In this appendix we calculate holographically the glueball spectra in $d$ dimensions using the soft wall model in $(d+1)$ dimensions. In addition, we use this spectra to fix the dilaton constant $ k $ in the 4-dimensional soft wall model from the glueball masses in $ (2+1) $ dimensions obtained from lattice \cite{Athenodorou:2016ebg}.

The soft wall action for a massive scalar field in $ AdS_{d+1} $, up to dimensional parameters, reads
\begin{equation}
S =  \int d^{d+1} x \sqrt{-g} e^{-\Phi(z)}\left[g^{MN} \partial_M{\cal G}\partial_N{\cal G} + M^2_{d+1} {\cal G}^2\right],
\end{equation}
where $ \Phi(z)=kz^2 $ is the dilaton field and $ \mathcal{G} = \mathcal{G}(z,x^{\mu}) $ is a scalar field with mass $ M_{d+1} $ in $ AdS_{d+1} $. The corresponding equations of motion coming from this action are given by
\begin{equation}
\partial_M\left[ \sqrt{-g} \;  e^{-\Phi(z)} g^{MN} \partial_N {\cal G}\right] - \sqrt{-g} e^{-\Phi(z)} M^2_{d+1} {\cal G} = 0\,.
\end{equation}
Using the Ansatz $ \mathcal{G}(z,x^{\mu}) = g(z)e^{-iP_{\mu}x^\mu}$ with the metric given by
\begin{equation}
ds^2 = \dfrac{L^2}{z^2}(dz^2+\eta_{\mu\nu}dx^{\mu}dx^{\nu}),
\end{equation} 
we obtain the following second-order differential equation for $ g(z) $
\begin{equation}
z^2g''(z) + \left(1-d-2kz^2\right)zg'(z) + \left(m^2 z^2-(M_{d+1}L)^2\right)g(z) = 0.
\end{equation}
Up to some normalization constant and a global phase, the solution for $ g(z) $ which is regular at $ z\rightarrow0 $ and $ z\rightarrow\infty $ is given by
\begin{equation}
g(z) = 
(\sqrt{k}z)\,^{\Delta}\,_1F_1\left(-n;\left(\Delta-\frac{d}{2}+1\right);kz^2\right),
\end{equation}
where $ n $ is an integer, $\Delta = \frac{d}{2} + \sqrt{\frac{d^2}{4}+(M_{d+1}L)^2}$ and $ _1F_1\left(a;b;x\right) $ is the confluent hypergeometric function. The mass spectra is given by
\begin{equation}
m_{n}^2 = \left(4n + d + \sqrt{d^2+ 4(M_{d+1}L)^2}\right)k.
\end{equation}
This result is valid for $ k>0 $. However, for $ k<0 $, this result still remains unchanged, but in this case we have to put the absolute value of $ k $. Therefore, we have for any $k\neq0$
\begin{equation}
m_{n}^2 = \left(4n + d + \sqrt{d^2+ 4(M_{d+1}L)^2}\right)|k|.
\end{equation}
It is easy to see that for $ d=4 $ and $ M_{5}=0 $ one reproduces the result $ m^{2}_{n}=(4n+8)k $, which is the spectrum of scalar glueballs in $ (3+1) $ dimensions found in \cite{Colangelo:2007pt}. The integer $ n $ is the radial quantum number, with $ n = 0 $ being the ground state and $ n = 1,2... $ being the radial excitations.

In our particular case, for $ d=3 $ and $ M_{4}=0 $, we have
\begin{equation}
m_{n}^2 = \left(4n + 6\right)k.
\end{equation}
Using the masses \eqref{Masses_Lattice} for the lightest glueball in $ (2+1) $ dimensions from the lattice \cite{Athenodorou:2016ebg} and setting $ n=0 $, we can fix the dilaton constant $ k $ as
\begin{equation}
\dfrac{k}{\sigma}  = \left\{\begin{array}{ll} 3.74 \qquad SU(2), \\ 3.18 \qquad SU(3), \\
2.82 \qquad SU(N\rightarrow\infty),
\end{array}\right.
\end{equation}
in units of the string tension squared.


\begin{thebibliography}{ABC}
\bibitem{comment1}
Since in the next sections we will be interested in computing on-shell Euclidean actions, we have decided to start this section by continuing to Euclidean signature with a compact time direction, $0\leq\tau\leq\beta$, where $ \beta = \frac{1}{T} $.
\end{thebibliography}
 \end{document}